\documentclass[11pt,a4paper]{article}
\usepackage{jcappub}
\usepackage{float}
\usepackage{amssymb}
\usepackage{graphicx}
\usepackage{amsmath}
\usepackage{hyperref}
\usepackage{amsfonts}
\usepackage{subfigure}

\title{Constraints on perturbative \\$f(R)$ gravity via neutron stars}

\author[a]{Sava\c{s} Arapo\u{g}lu,} 
\author[b]{Cemsinan Deliduman}
\author[a]{and K. Yavuz Ek\c{s}i}

\affiliation[a]{\.{I}stanbul Technical University, Faculty of
Science and Letters, Physics Engineering Department,\\  
Maslak 34469, \.{I}stanbul, Turkey}
\affiliation[b]{Mimar Sinan Fine Arts University, Department of Physics,\\ 
Be\c{s}ikta\c{s} 34349, \.{I}stanbul, Turkey}

\emailAdd{arapoglu@itu.edu.tr}
\emailAdd{cemsinan@msgsu.edu.tr}
\emailAdd{eksi@itu.edu.tr}

\abstract{We study the structure of neutron stars in perturbative $f(R)$
gravity models with realistic equations of state.  We obtain
mass--radius relations in a gravity model of the form $f(R)=R+\alpha
R^2$. We find that deviations from the 
results of general relativity, comparable to the variations due to using different
equations of state (EoS'), are induced 
for $|\alpha| \sim 10^{9}$ cm$^{2}$. Some of the soft EoS' that are excluded within the framework of general relativity
can be reconciled with the 2 solar mass neutron star recently observed for certain values of $\alpha$ within this range. 
For some of the EoS' we find that a new solution branch, which allows highly massive neutron stars, exists for values of $\alpha$ greater than a few $10^9$ cm$^2$. We find constraints on $\alpha$ for a variety
of EoS' using the recent observational
constraints on the mass--radius relation.
These are all 5 orders of magnitude smaller than the recent constraint
obtained via Gravity Probe B for this gravity model. 
The associated length scale
$\sqrt{\alpha}\sim 10^5$ cm is only an order of magnitude smaller
than the typical radius of a neutron star, the probe used in this
test. This implies that real deviations from general relativity can be even smaller.}

\keywords{modified gravity, neutron stars}

\begin{document}

\maketitle

\section{Introduction}

The current accelerated expansion of the universe has been confirmed
by many independent observations. The supporting evidence comes from
the supernovae Ia data \cite{Perlmutter, Riess1, Riess2}, cosmic
microwave background radiation \cite{Spergel1, Spergel2, Komatsu},
and the large scale structure of the universe \cite{Tegmark1,
Tegmark2}. Although the cosmological constant is arguably the
simplest explanation and the best fit to all observational data, its
theoretical value predicted by quantum field theory is many orders
of magnitude greater than the value to explain the current
acceleration of the universe.  This problematic nature of
cosmological constant has motivated an intense research for
alternative explanations and the reasonable approaches in this
direction can be divided into two main categories, both of them
introducing new degrees of freedom \cite{Uzan}: The first approach
is to add some unknown energy-momentum component to the right hand
side of Einstein's equations with an equation of state $p/\rho
\approx -1$, dubbed \textit{dark energy}.  In the more radical
second approach, the idea is to modify the left hand side of
Einstein's equations, so-called \textit{modified gravity}.  Trying
to explain such perplexing observations by modifying gravity rather
than postulating an unknown dark energy has been an active research
area in the last few years and in this paper we adopt this path.

A modified theory of gravity has to explain the late time cosmology,
and also be compatible with the constraints obtained from solar
system and laboratory tests. However, it is not easy to construct
theories of gravity with these requirements.  Nevertheless, a class
of theories, called $f(R)$ models \cite{Odintsov-rev, Sotiriou-rev, deFelice-rev},
has attracted serious attention possibly because of its (deceptive)
simplicity.  Today there exist viable $f(R)$ models which are
constructed carefully to be free of instabilities, and to pass the
current solar system and laboratory tests \cite{Nojiri1, Nojiri2, Nojiri3, Cognola, Hu-Sawicki,
Appleby-Battye, Starobinsky, Miranda}.

The strong gravity regime \cite{Dimitri-rev} of these theories is
another way of checking their viability.  In this regime,
divergences stemming from the functional form of f(R) may prevent
the existence of relativistic stars in such theories \cite{Briscese,
Abdalla, Nojiri4, Bamba, Kobayashi-Maeda, Frolov, Nojiri5}, but
thanks to the chameleon mechanism the possible problems jeopardizing
the existence of these objects may be avoided \cite{Tsujikawa,
Upadhye-Hu}.  Furthermore, there are also numerical solutions
corresponding to static star configurations with a strong
gravitational field \cite{Babichev1, Babichev2} where the choice of
the equation of state for the star is crucial for the existence of
solutions, and therefore a polytropic equation of state is used in
these works to overcome the possible problems related to the
equation of state.

Another approach to probe the viability of f(R) theories in the
strong gravity regime is to use a method called perturbative
constraints, or order reduction \cite{Eliezer, Jaen}.  The
motivation behind using this technique in the present context is the
thought that the reason of all the problems encountered in modifying
gravity may be the outcome of considering these modifications as
exact ones.  The main issue with the exact modifications are the
problems arising in curvature scales which are not originally aimed
by these modifications.  In the perturbative constraints approach,
the modifications are viewed as next to leading order terms to the
terms coming from Einstein's General Relativity.  Treating f(R)
gravity via perturbative constraints at cosmological scales is
considered in \cite{DeDeo, Cooney1} and for compact objects in
\cite{Cooney2}.  In this manuscript we further examine the existence
and properties of relativistic stars in the context of f(R) models
via perturbative constraints.

Specifically, we study a $f(R)$ model of the form $f(R)=R+\alpha
R^2$ and constrain the value of $\alpha$ with the recent constraints
on the mass-radius relation \cite{OBG10}. Such a gravity model in
the weak field limit is known to reduce to Yukawa-like potentials
and has been recently constrained by binary pulsar data \cite{naf}
as $\alpha \lesssim 5 \times 10^{15}$ cm$^2$. Here we find that in
the strong gravity regime the constraint on perturbative parameter
is $\alpha \lesssim 10^{10}$ cm$^2$. This value does not contrast
with the value obtained in \cite{naf} as they argue that the value
of $\alpha$ could be different at different length scales.

The plan of the paper is as follows:  In section II, we assume a
perturbative form of $f(R)$ modified gravity model and obtain the
field equations. Assuming also perturbative forms of metric
functions and the hydrodynamical quantities we obtain the modified
Tolman--Oppenheimer--Volkoff (TOV) equations. In section III, the
modified TOV equations are solved numerically for various forms of
the equation of state, the functional form of $f(R)$, and various
values of the perturbation parameter $\alpha$. Finally, in the
discussion section, we comment on the results of numerical study and
on the significance of the scale of the perturbation parameter
$\alpha$.

\section{Modified TOV equations of $f(R)$ gravity}

The action of $f(R)$ gravity models is the simplest generalization
of the Einstein--Hilbert action. Here, instead of having a linear
function of Ricci scalar as the Lagrangian density we have a
function of it:
\begin{equation}\label{action}
S=\frac{1}{16\pi}\int d^4x \sqrt{-g}f(R) + S_{{\rm matter}}\quad ,
\end{equation}
where $g$ denotes the determinant of the metric $g_{\mu\nu}$, and
$R$ is the Ricci scalar. We set $G=1$ and $c=1$ in the action and in
the rest of this section. Here we are considering the metric
formalism of gravity and therefore matter only couples to the
metric, and the Levi--Civita connection is derived from the
metric.

A straightforward variation of the action (\ref{action}) with
respect to the metric gives fourth order differential equations of
$g_{\mu\nu}$. However, treating higher than second order
differential equations in 4 dimensions is problematic. For this
reason, we adopt a perturbative approach as suggested in
\cite{Cooney2} and choose $f(R)$ such that all terms higher than
second order will be multiplied by a small parameter $\alpha$.
The meaning of smallness of the parameter $\alpha$ is explained in the next section.

Since this is a perturbative approach, the action
and the field equations must have the form of those of general
relativity for $\alpha=0$. So we choose the function $f(R)$ to have the form
\begin{equation}\label{fR}
    f(R)=R+\alpha h(R)+\mathcal{O}(\alpha ^{2})
\end{equation}
without a constant piece, i.e. without a cosmological constant. Here
$h(R)$ is, for now, an arbitrary function of $R$ and
$\mathcal{O}(\alpha ^{2})$ denotes the possible higher order
corrections in $\alpha$. Variation of the action (\ref{action}) with
respect to the metric, with the form of $f(R)$ as given in (\ref{fR}),
results in field equations which are
\begin{equation}\label{field}
(1+\alpha h_{R})G_{\mu \nu }-\frac{1}{2}\alpha(h-h_{R}R)g_{\mu \nu
}-\alpha (\nabla _{\mu }\nabla _{\nu }-g_{\mu \nu }\Box )h_{R}=8\pi
T_{\mu \nu }
\end{equation}
where $G_{\mu\nu}=R_{\mu\nu}-\frac12Rg_{\mu\nu}$ is the Einstein
tensor and $h_R=\frac{dh}{dR}$ is the derivative of $h(R)$ with
respect to the Ricci scalar.

We are interested in spherically symmetric solutions of these field
equations inside a neutron star, so we choose a spherically
symmetric metric with two unknown, independent functions of $r$,
\begin{equation}\label{metric}
    ds^2= -e^{2\phi_\alpha}dt^2 +e^{2\lambda_\alpha}dr^2 +r^2 (d\theta^2
    +\sin^2\theta d\phi^2).
\end{equation}

Perturbative solution of the field equations means that
$g_{\mu\nu}$ can be expanded perturbatively in $\alpha$ and
therefore the metric functions have the expansions $\phi_\alpha =
\phi +\alpha\phi_1 + \ldots$ and $\lambda_\alpha = \lambda
+\alpha\lambda_1 + \ldots$. The energy--momentum tensor, which is
on the right hand side of the field equations, is the
energy-momentum tensor of the perfect fluid. For a perturbative
solution, then, hydrodynamics quantities are also defined
perturbatively: $\rho_\alpha = \rho +\alpha\rho_1 + \ldots$ and
$P_\alpha = P +\alpha P_1 + \ldots$. Note that we denote zeroth order
quantities, which can be obtained solving Einstein
equations, without a subscript.

Then the ``tt'' and ``rr'' components of field equations become
\begin{eqnarray}
  -8\pi \rho_\alpha &=& -r^{-2} +e^{-2\lambda_\alpha}(1-2r\lambda_\alpha')r^{-2}
                +\alpha h_R(-r^{-2} +e^{-2\lambda}(1-2r\lambda')r^{-2}) \nonumber \\
             && -\frac12\alpha(h-h_{R}R) +e^{-2\lambda}\alpha[h_R'r^{-1}(2-r\lambda')+h_R''] \label{f-tt},\\
  8\pi P_\alpha &=& -r^{-2} +e^{-2\lambda_\alpha}(1+2r\phi_\alpha')r^{-2}
                +\alpha h_R(-r^{-2} +e^{-2\lambda}(1+2r\phi')r^{-2}) \nonumber \\
             && -\frac12\alpha(h-h_{R}R) +e^{-2\lambda}\alpha h_R'r^{-1}(2+r\phi'), \label{f-rr}
\end{eqnarray}
where prime denotes derivative with respect to radial distance, $r$.

We would like to solve $\rho_\alpha$ and $P_\alpha$ up to order
$\alpha$. On the right hand side, terms containing $h_R$ and its
derivatives are already first order in $\alpha$. Therefore, for the
quantities that are multiplied by them, we use the zeroth order
quantities which can be obtained from the Einstein equations written
for this metric. In the equations above, hence, in terms that are
multiplied by $\alpha$ we have already written zeroth order
quantities. Using the ``tt" and ``rr" components of Einstein
equations,
\begin{eqnarray}
  -8\pi \rho &=& -r^{-2} +e^{-2\lambda}(1-2r\lambda')r^{-2} \label{e-tt},\\
  8\pi P &=& -r^{-2} +e^{-2\lambda}(1+2r\phi')r^{-2}, \label{e-rr}
\end{eqnarray}
we replace terms multiplied by $h_R$ in (\ref{f-tt})
and (\ref{f-rr}) with $-8\pi \rho$ and $8\pi P$, respectively.
Additionally, combining $\alpha$ order terms on the left hand sides of
(\ref{f-tt}) and (\ref{f-rr}) we obtain,
\begin{eqnarray}
  8\pi r^{2} \rho_\alpha &=& 1-e^{-2\lambda_\alpha}(1-2r\lambda_\alpha') \nonumber \\
            && +\alpha h_R r^{2}\left[ 8\pi \rho  +\frac12\left(\frac{h}{h_R}-R\right)
                -e^{-2\lambda}\left(r^{-1}(2-r\lambda')\frac{h_R'}{h_R}+\frac{h_R''}{h_R}\right)\right] \label{f-tt2},\\
  8\pi r^{2} P_\alpha &=& -1+e^{-2\lambda_\alpha}(1+2r\phi_\alpha') \nonumber \\
            && +\alpha h_R r^{2}\left[ 8\pi P -\frac12\left(\frac{h}{h_R}-R\right)
                +e^{-2\lambda}r^{-1}(2+r\phi')\frac{h_R'}{h_R}\right]. \label{f-rr2}
\end{eqnarray}

To define a mass parameter, we assume a solution that has the same
form of the exterior solution for the metric function
$\lambda_\alpha$. This form of solution has been previously suggested by
the authors of \cite{Cooney2}. Therefore we define
\begin{equation}\label{mass}
    e^{-2\lambda_\alpha}=1-\frac{M_\alpha}{r}.
\end{equation}
Here, similar to $\rho_\alpha$, $M_\alpha$ is expanded in $\alpha$ as
$M_\alpha = M +\alpha M_1 + \ldots$, where $M$ is the zeroth order
solution, which, in general relativity, is given in terms of $\rho$ as
\begin{equation}\label{mass_GR}
    M=8\pi \int \rho(r) r^2 dr.
\end{equation}
Taking the derivative of $M_\alpha$ with respect to $r$ one obtains
\begin{equation}\label{dMa/dr}
    \frac{dM_\alpha}{dr}= 1-e^{-2\lambda_\alpha}(1-2r\lambda_\alpha').
\end{equation}
Substituting this into (\ref{f-tt2}) and arranging terms, one gets
the first modified TOV equation,
\begin{eqnarray}
  \frac{dM_\alpha}{dr} &=& 8\pi r^{2} \rho_\alpha -\alpha h_R \left[
  \begin{array}{l}
    8\pi r^{2}\rho  +\frac{r^2}{2} (\frac{h}{h_R}-R) \\
    +(4\pi\rho r^3-2r+\frac32 M)\frac{h_R'}{h_R}-r(r-M)\frac{h_R''}{h_R}
  \end{array}
    \right] \label{1stTOV}.
\end{eqnarray}
To obtain this equation, we also substitute general relativistic form of
(\ref{mass}), $e^{-2\lambda}=1-\frac{M}{r}$.

The conservation equation of energy-momentum tensor of a perfect
fluid, $\nabla^\mu T_{\mu\nu}=0$, is equivalent to the hydrostatic
equilibrium equation,
\begin{equation}\label{hydro}
    \frac{dP_\alpha}{dr}=-(\rho_\alpha
    +P_\alpha)\frac{d\phi_\alpha}{dr}.
\end{equation}
Therefore in order to get the second modified TOV equation we pull
$\frac{d\phi_\alpha}{dr}$ from the ``rr'' field equation, eq.(\ref{f-rr2}). After some straightforward manipulations one gets,
\begin{eqnarray}
  2(r-M_\alpha)\frac{d\phi_\alpha}{dr} &=& 8\pi r^{2}P_\alpha + \frac{M_\alpha}{r}
  -\alpha h_R \left[
  \begin{array}{l}
    8\pi r^{2}P  +\frac{r^2}{2} (\frac{h}{h_R}-R) \\
    +(2r-\frac32M+4\pi Pr^3)\frac{h_R'}{h_R}
  \end{array}
    \right] \label{2ndTOV}.
\end{eqnarray}

Note that when one sets $\alpha$ to zero, one gets the original TOV
equations for general relativistic quantities. Similar to the case
in general relativity, the modified TOV equations, (\ref{1stTOV}),
(\ref{hydro}) and (\ref{2ndTOV}), are solved numerically for some
special functional form of $h(R)$. Obviously perturbation expansion
parameter $\alpha$ introduces a new scale into the theory. By
choosing a realistic equation of state we compute mass--radius relation for various
values of $\alpha$ and this way we put a bound on $\alpha$ for
perturbative $f(R)$ gravity models with various forms of $h(R)$.\footnote{A realistic equation
of state employs different physics for different regions of the
neutron star, as opposed to a single polytropic relation prevailing
throughout the star.}
This numerical analysis is explained in the next section.

\section{Numerical model and astrophysical constraints on the value of $\alpha$}

Eqs. (\ref{1stTOV}) and (\ref{2ndTOV}) describe any spherical mass
distribution in a general $f(R)$ theory in perturbative approach. Interesting results would
be obtained in the case of neutron stars which have the highest
compactness ratio $\eta=2GM_{\ast}/c^2R_{\ast}$ and curvatures
$\xi=GM_{\ast}/c^2R_{\ast}^3$ \cite{Dimitri-rev}. In order to specialize these
equations for describing neutron stars they must be supplemented by
an appropriate equation of state (EoS). However, the EoS of
nuclear matter at the densities prevailing in neutron stars is not
very well constrained by nuclear scattering data and there is a
number of EoS leading to different mass-radius (M-R)
relations for neutron stars.

For integrating the TOV equations in GR, it is
possible to interpolate the tabulated relation between density and
pressure. In $f(R)$ theories, where one needs high order derivatives
of pressure with respect to density, interpolation leads to
numerical problems. In order to circumvent this problem, we
employ analytical expressions obtained by fitting the tabulated data. 
Such analytical representations are already provided by \cite{HP04} for two EoS', FPS and SLy.    
For the rest of the EoS', namely AP4, GS1, MPA1, and MS1, we used
analytical representations provided by \cite{GE11} obtained by fitting the tabulated data with
a function which is an extension of the 
function provided in  \cite{HP04}. The 6 EoS' we have chosen constitute a representative sample 
(see figure 1 in \cite{OBG10}) for multi-nucleonic and condensate inner composition possibilities 
(see \cite{LP01} for the description of all these EoS'). 
We have not employed any strange quark matter EoS. As such stars are not gravitationally bound, 
alternative gravity models does not produce different mass-radius relations for such objects.

We numerically integrate Eqs. (\ref{1stTOV}),
(\ref{hydro}) and (\ref{2ndTOV}) supplemented by the analytical
expression for the EoS, employing a
Runge-Kutta scheme with fixed step size of $\Delta r=0.001$ km. We obtain
a sequence of equilibrium configurations by varying the central
density $\rho_c$ from $2\times 10^{14}$ g cm$^{-3}$ to $1\times
10^{16}$ g cm$^{-3}$ (to $2\times
10^{16}$ g cm$^{-3}$ in some cases) in 200 logarithmically equal steps. This traces
a mass-radius relation for a certain equation of state. We then repeat this
procedure for a range of $\alpha$ to see the effect of the higher
order terms in perturbative $f(R)$ gravity.

Recently, the authors of \cite{OP09} showed that the measurement of
masses and radii of three neutron stars are sufficient for
constraining the pressure of nuclear matter at densities a few times
the density of nuclear saturation. These data are provided by the
measurements on the neutron stars EXO 1745-248 \cite{OGP09}, 4U
1608-52 \cite{GOCW09} and 4U 1820-30 \cite{GWCO10} by the methods
described in the cited papers. We use the constraints on the
M-R relation of neutron stars given in \cite{OBG10}, which
is a union of these three constraints.\footnote{For a critic of these constraints see 
\cite{ste10}.}  The constraint of \cite{OBG10} is shown in all M-R 
plots as the region bounded by the thin black line. Note that the M-R constraint 
in the first version of \cite{OBG10} is larger in the published version. In the earlier versions 
of our work we employed the earlier tighter constraint and reached somewhat different conclusions for FPS and SLy EoS.

Apart from the above constraint, the recent measurement \cite{dem10} 
of the mass of the neutron star 
PSR J1614-2230 with $1.97 \pm 0.04\, M_{\odot}$ provides a stringent constraint on any
M-R relation that can be obtained with a combination of $\alpha$ and EoS. 
This constraint is shown as the horizontal black line with its error shown in grey.
Any viable combination of $\alpha$ and EoS must yield a M-R relation with 
a maximum mass exceeding this measured mass.

The constraints on the M-R relation obtained by
\cite{OBG10} and the 2 solar mass neutron star PSR J1614-2230 
exclude many of the possible EoS' if one assumes GR as 
the ultimate classical theory of gravity. 
In the gravity model employed here, the value of $\alpha$ provides a new 
degree of freedom such that 
some of the EoS', which are excluded within the framework of GR, 
can now be reconciled with the observations for certain values of $\alpha$.
In the following we discuss this for all EoS' individually. To save space in the figures,
we define the parameter $\alpha_9 \equiv \alpha/10^9$ cm$^2$. 
We show the stable configurations ($dM/d\rho_c > 0$) with solid lines and the unstable configurations with dashed lines of the same color.

\begin{figure}[H]
\begin{center}
\includegraphics{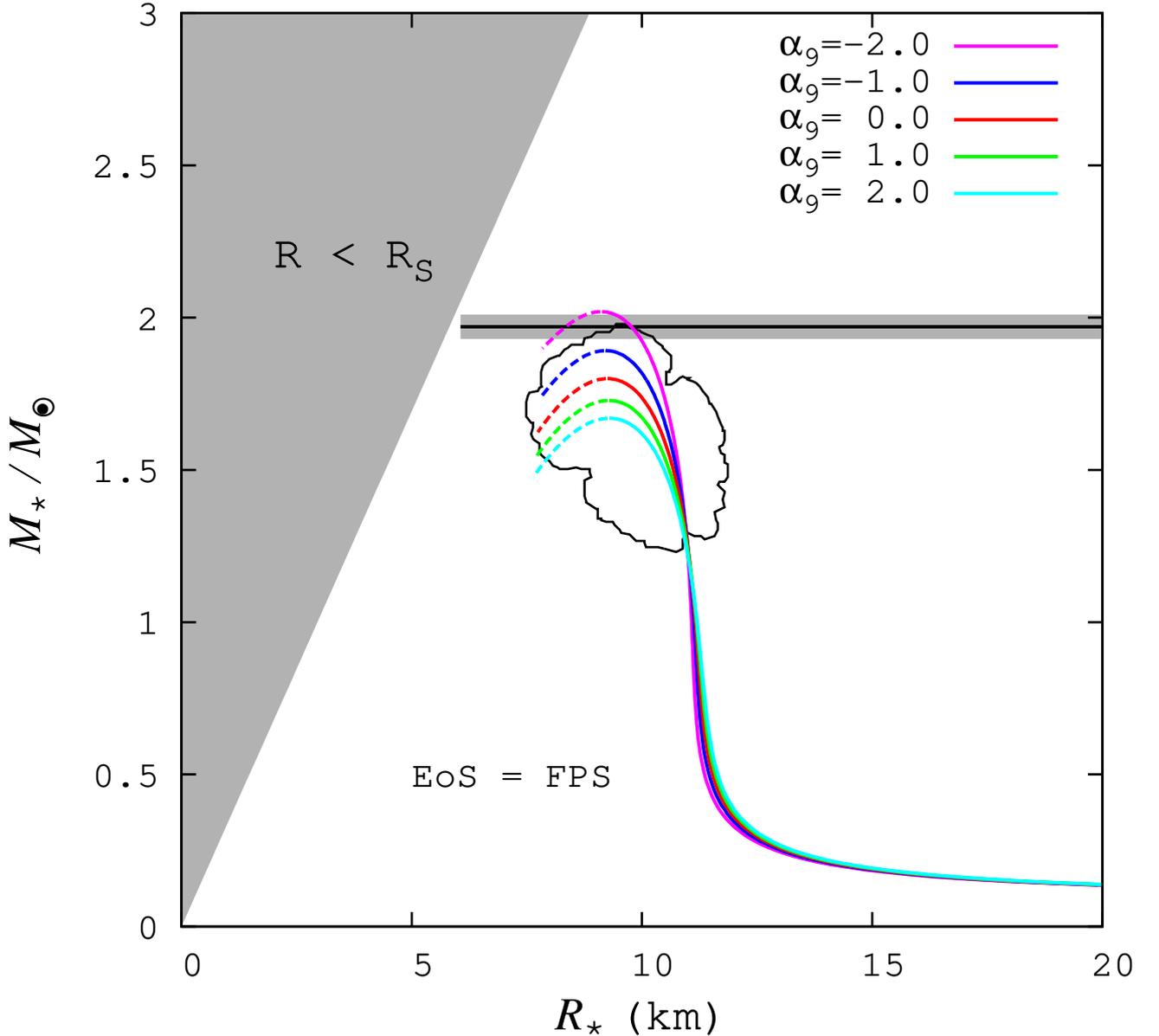}
\caption{\label{fig_FPS} M-R relation
obtained with $f(R)=R+\alpha R^2$
using the FPS. The observational
constraints of \cite{OBG10} is shown with the thin black contour; the measured mass $M=1.97 \pm 0.04\, M_{\odot}$ of PSR J1614-2230 \cite{dem10} is shown as the horizontal black line with grey errorbar. Each solid line corresponds to a stable configuration for a specific value of $\alpha$. Dashed lines show the solutions for unstable configurations ($dM/d\rho_c < 0$). The grey shaded region shows where the total mass would be enclosed within its Schwarzschild radius. The red line ($\alpha=0$) shows the
result for GR. $M_{\max}$ and $R_{\min}$ increase for decreasing values of $\alpha$. Variations in the M-R relation comparable to employing different EoS' can be obtained 
for $|\alpha| \sim 10^{9}$ cm$^{2}$. Using $\alpha \lesssim 10^{8}$ cm$^{2}$ gives M-R relations that can not be distinguished from the GR result on this plot. 
}
\end{center}
\end{figure}

\begin{figure}[H]
\begin{center}
\includegraphics{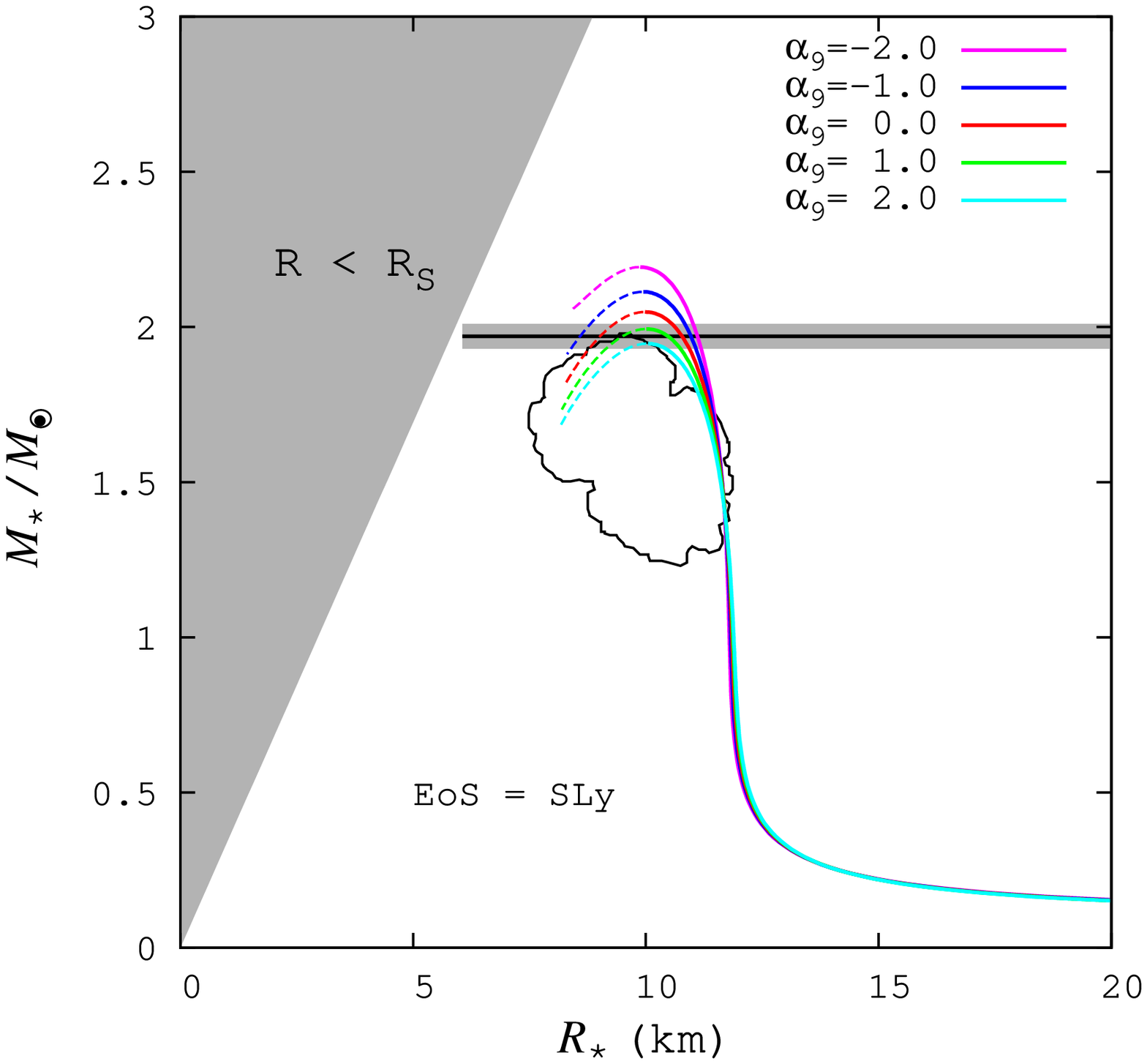}
\caption{\label{fig_SLY} M-R relation for the SLy. The notation in the figure is the same as 
that of Figure~\ref{fig_FPS} and the results are discussed in the text.}
\end{center}
\end{figure}

\begin{figure}[H]
\begin{center}
\includegraphics{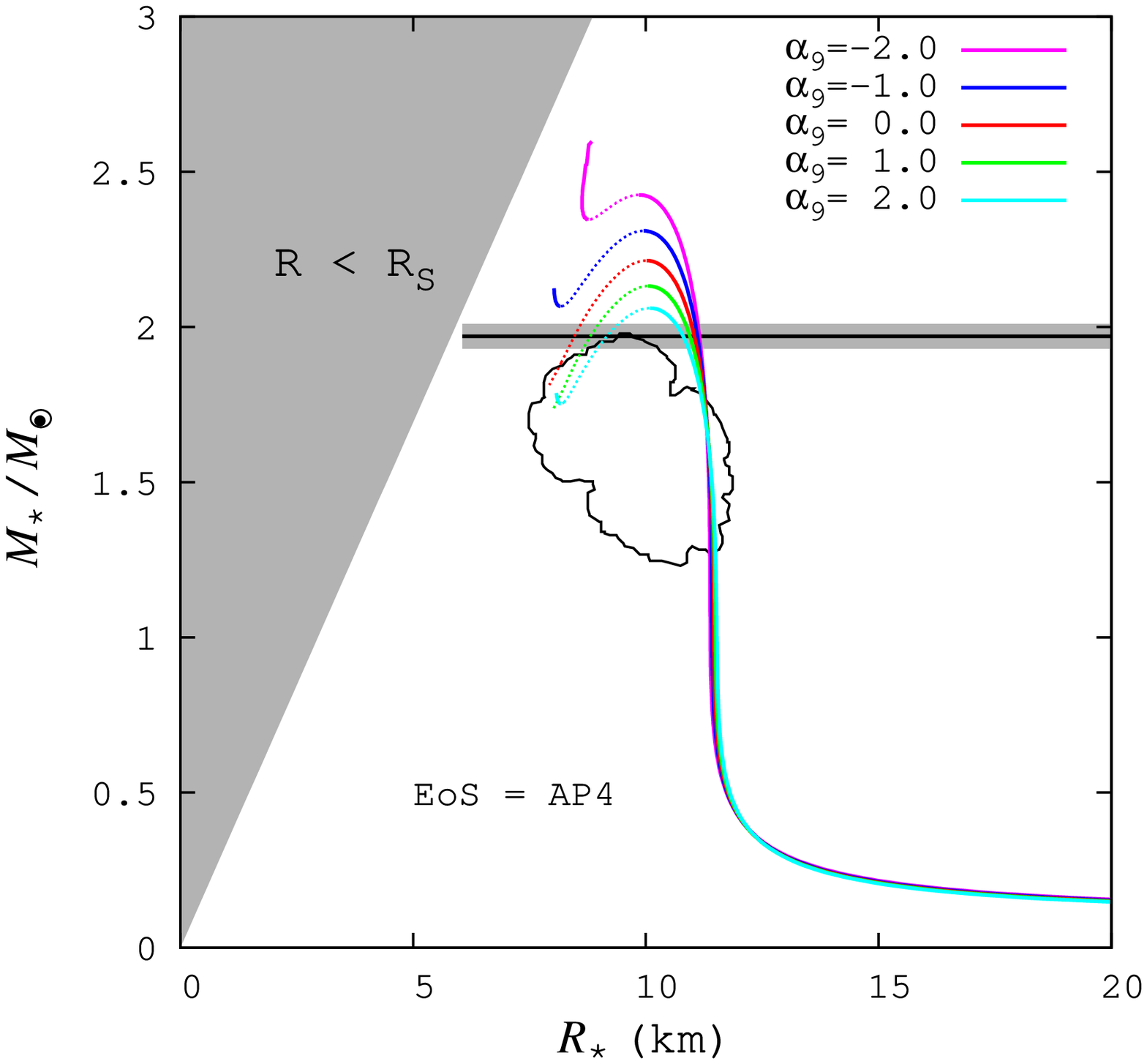}
\caption{\label{fig_AP4} M-R relation for the AP4. The notation in the figure is the same as 
that of Figure~\ref{fig_FPS} and the results are discussed in the text.}
\end{center}
\end{figure}

\begin{figure}[H]
\begin{center}
\includegraphics{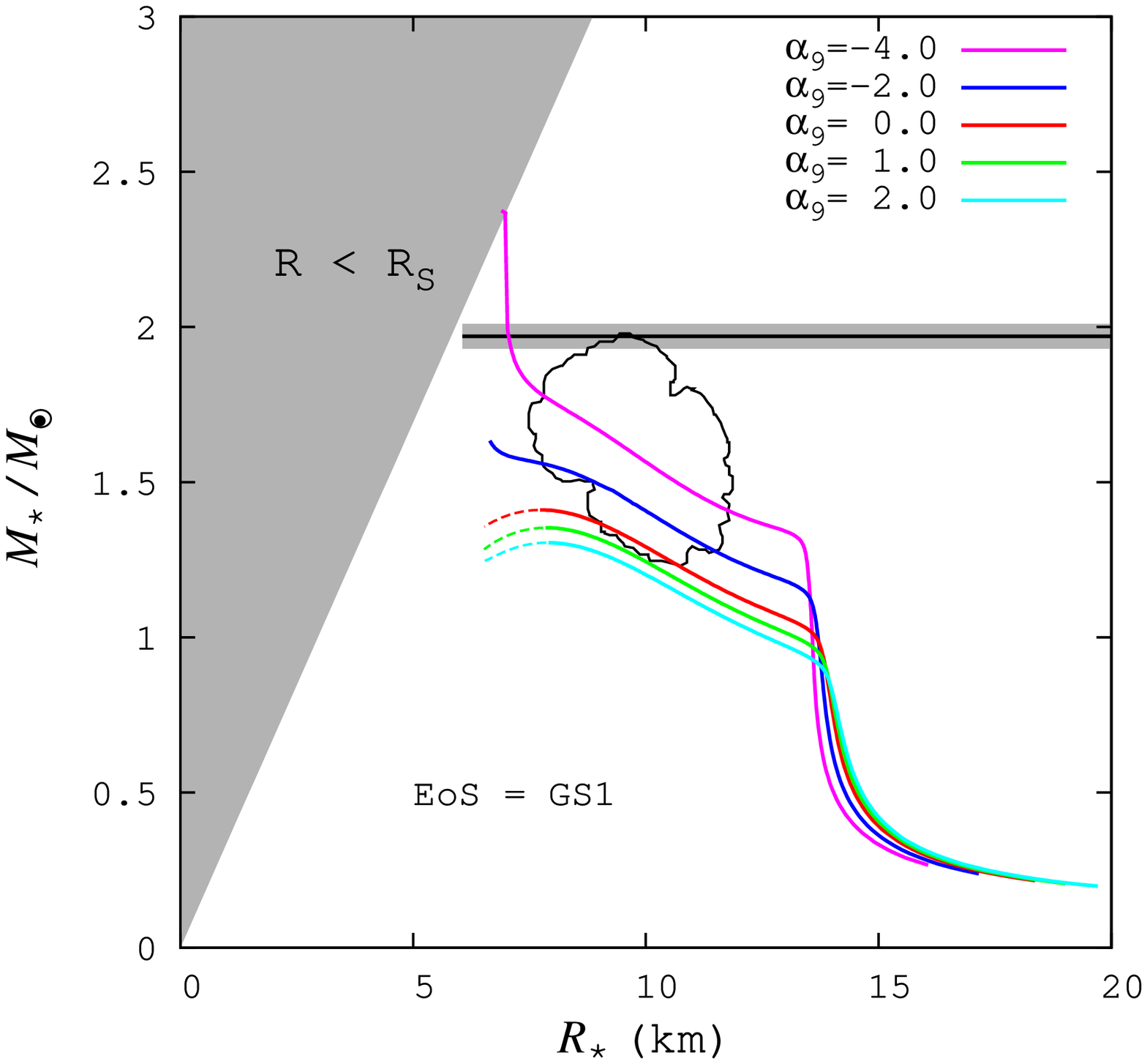}
\caption{\label{fig_GS1} M-R relation for the GS1. The notation in the figure is the same as 
that of Figure~\ref{fig_FPS} and the results are discussed in the text.}
\end{center}
\end{figure}

\begin{figure}[H]
\begin{center}
\includegraphics{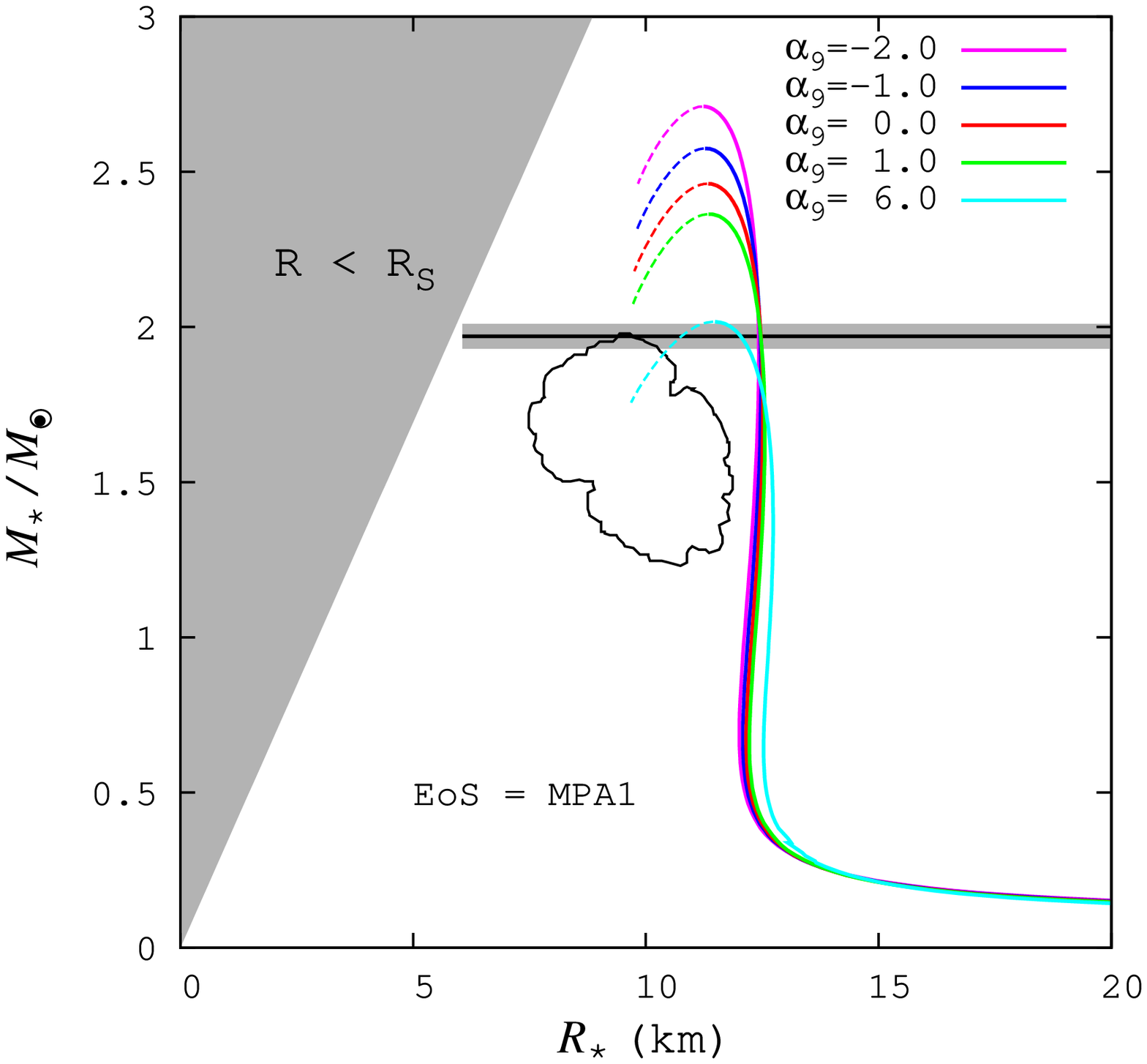}
\caption{\label{fig_MPA1} M-R relation for the MPA1. The notation in the figure is the same as 
that of Figure~\ref{fig_FPS} and the results are discussed in the text.}
\end{center}
\end{figure}

\begin{figure}[H]
\begin{center}
\includegraphics{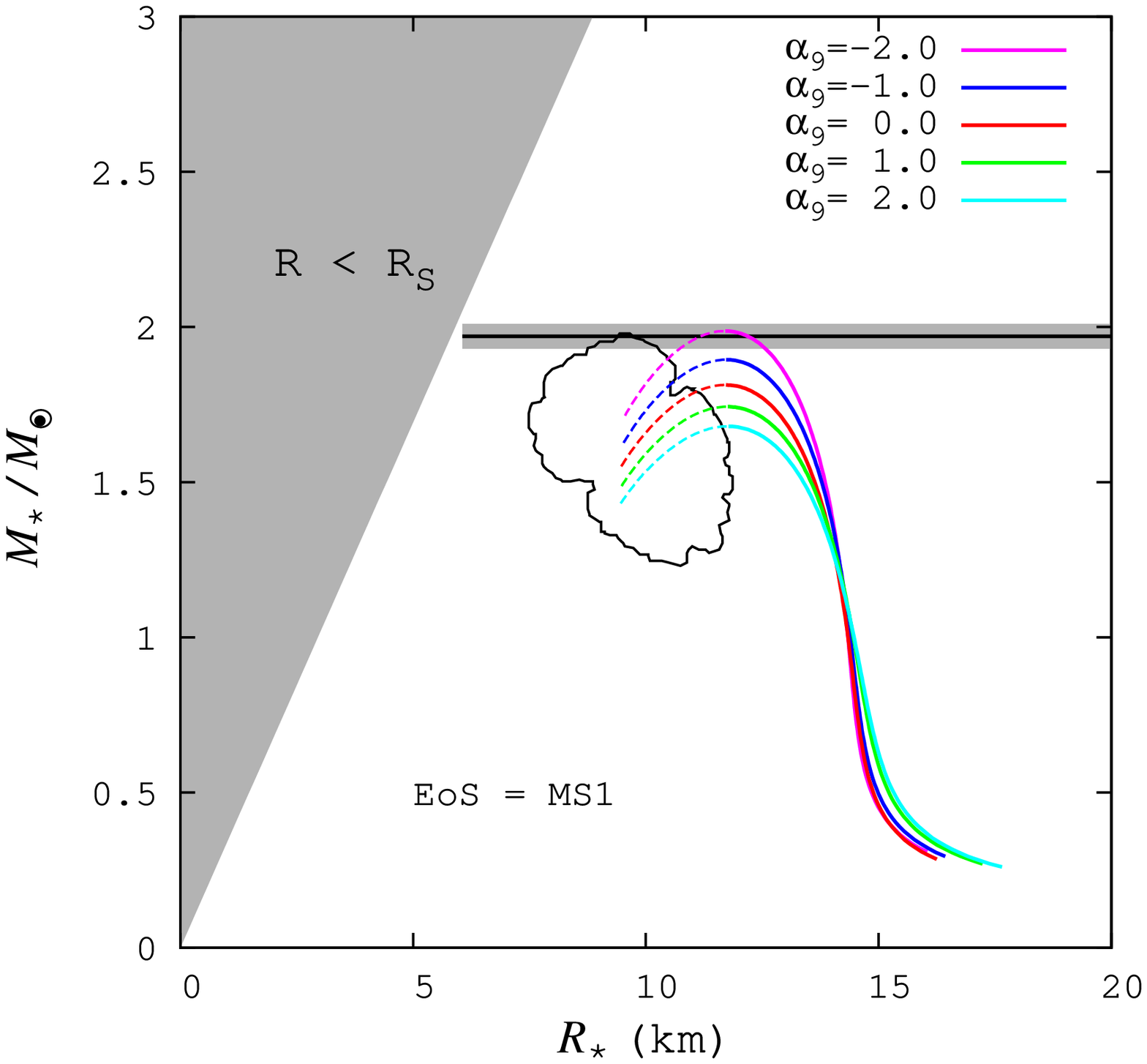}
\caption{\label{fig_MS1} M-R relation for the MS1. The notation in the figure is the same as 
that of Figure~\ref{fig_FPS} and the results are discussed in the text.}
\end{center}
\end{figure}

In figures, we show our results (mass versus radius, M-R, relations) for 6 representative EoS'
for the $f(R)=R+\alpha R^2$ gravity model. Results for each EoS are summarized as follows:

\begin{itemize}

\item {\bf FPS (Figure \ref{fig_FPS}):}  
For FPS \cite{ref_FPS}, the maximum mass that a neutron star can have, within GR, 
is about $1.8\,M_{\odot}$ and is less than the measured mass of PSR J1614-2230. 
This means FPS can not represent the EoS of neutron stars in GR ($\alpha =0$).
The maximum mass increases 
with decreasing value of $\alpha$ and we find that,
for $\alpha_9<-2$,  the maximum mass
becomes $M_{\max}=2.04 M_{\odot}$.   
We thus find that FPS is consistent with the measurement of the maximum mass for $\alpha<-2 \times 10^9$ cm$^2$.
Nevertheless, this does not mean $\alpha= -10 \times 10^{9}$ cm$^2$ will be satisfying both constraints.
In fact for such large values of $|\alpha|$ we can not obtain M-R relations resembling known properties of neutron stars. 
Furthermore for $|\alpha|>10^{11}$ cm$^2$
the validity of the perturbative approach is dubious, as we mention in the discussion section.

\item {\bf SLy (Figure~\ref{fig_SLY}):}
SLy \cite{ref_SLY}
is consistent with both constraints within the framework of GR. For 
$\alpha>2\times 10^9$ cm$^2$, however, we see that $M_{\max}$ is less than 
the measured mass of PSR J1614-2230. We conclude for the gravity model employed here, 
$f(R)=R+\alpha R^2$, that SLy is consistent with the observations only if $\alpha<2\times 10^9$ cm$^2$.

\item {\bf AP4 (Figure~\ref{fig_AP4}):}
AP4 \cite{ref_AP4} is consistent with the constraints
as long as $\alpha<4\times 10^9$ cm$^2$. Interestingly, we find that a new stable solution branch, 
for which $dM/d\rho_c>0$ is satisfied, for values of $\alpha$ different than zero. This stable branch is obtained, for $\alpha_9=-2$ starting from central densities $8.6 \times 10^{15}$ g cm$^{-3}$.

\item {\bf GS1 (Figure~\ref{fig_GS1}):}
For GS1 \cite{ref_GS1}, the maximum mass in GR remains well below the measured mass
of PSR J1614-2230. The maximum mass of neutron stars for this EoS can reach 
up to $\sim 2\,M_{\odot}$ for $\alpha_9=-4$.  Starting from $\alpha_9=-2$ the stability condition ($dM/d\rho_c > 0$) is satisfied for the whole range of central densities considered.

\item {\bf MPA1 (Figure~\ref{fig_MPA1}):}
MPA1 \cite{ref_MPA1} provides a maximum mass 
above the observed mass of PSR J1614-2230 in GR, though it does not pass through the M-R constraint of \cite{OBG10}.
For $\alpha_9 > 6$ it can not satisfy the maximum mass constraint as well.

\item {\bf MS1 (Figure~\ref{fig_MS1}):}
The maximum mass for MS1 \cite{ref_MS1} satisfies the observed mass of PSR J1614-2230 only for $\alpha_9 <2$ though it moves away from the M-R constraint of \cite{OBG10} for such low values of $\alpha$.

\end{itemize}

\begin{figure}[H]
\begin{center}
\begin{minipage}[h]{0.49\linewidth}
\includegraphics[scale=0.45]{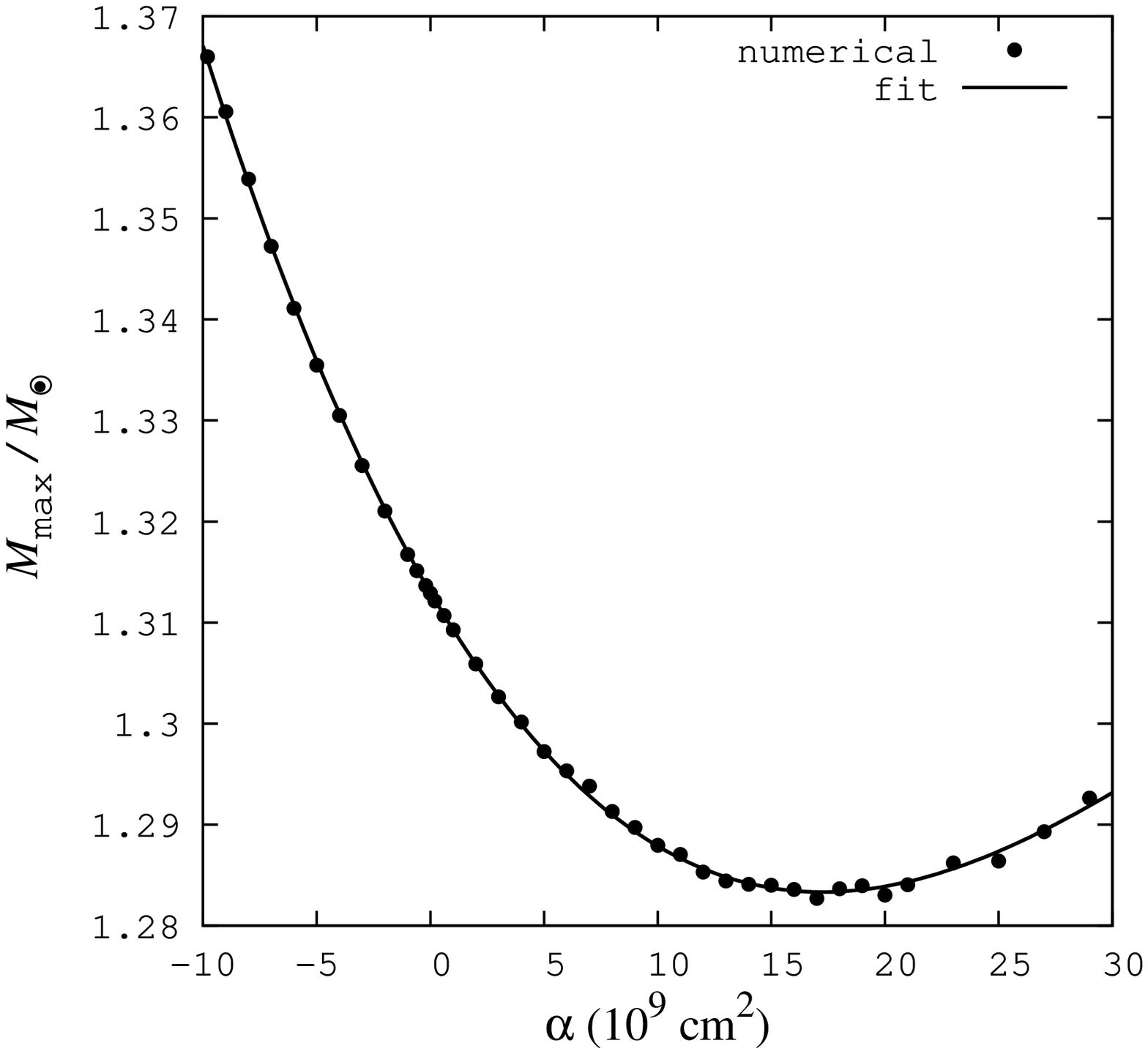}
\end{minipage}
\begin{minipage}[h]{0.49\linewidth}
\includegraphics[scale=0.45]{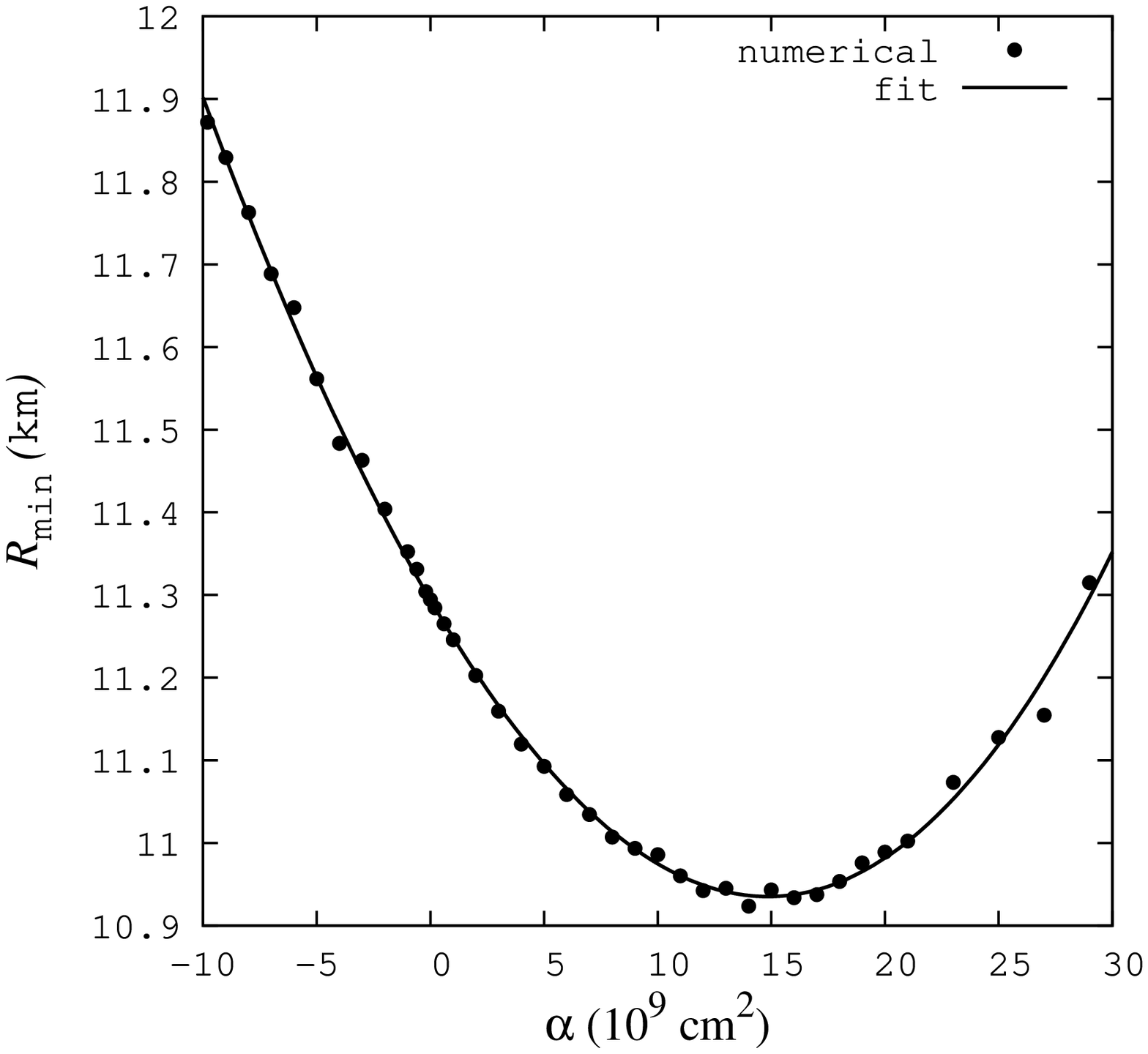}
\end{minipage}
\caption{$M_{\max}$ (left panel) and $R_{\min}$ (right panel) changing with $\alpha$ for the polytropic EoS given in Equation~(\ref{polytropic}).}
\label{fig_alpha}
\end{center}
\end{figure}

For all EoS' we observe that the maximum stable mass of a neutron star, $M_{\max}$, 
and its radius at this mass, $R_{\min}$, increases for decreasing values of $\alpha$, 
for the ranges we consider in the figures.  
There is no change in the behavior of $M_{\max}$ and $R_{\min}$ 
values while $\alpha$ changes sign. 
Thus the structure of neutron stars in GR ($\alpha=0$) does not 
constitute an extremal configuration in terms of $M_{\max}$ and $R_{\min}$.
In figure~\ref{fig_alpha} we show the dependence of these quantities
 on the value of $\alpha$ for the polytropic EoS
\begin{equation}
\rho =\left(\frac{P}{K}\right)^{1/\Gamma} +\frac{P}{\Gamma -1}
\label{polytropic}
\end{equation} 
used in \cite{Cooney2} where $\Gamma=9/5$ is the polytropic index and $K$ is a constant.
We fit the numerical results with cubic polynomials. For the maximum mass fitted with
\begin{equation}
M_{\max} = A \alpha_9^3 + B \alpha_9^2 + C \alpha_9 + M_0
\end{equation}
where $M_0$ is the maximum mass obtained for general relativity, we find that
$A= -1.30796\times 10^{-6}  \pm 5.547 \times 10^{-8} \, M_{\odot}$,
$B = 1.44851 \times 10^{-4} \pm 1.625 \times 10^{-6}\, M_{\odot}$ and
$C = -3.82907 \times 10^{-3}  \pm 1.234 \times 10^{-5} \, M_{\odot}$.
Interestingly, we find that for $\alpha \cong 15 \times 10^9$ cm$^2$ the maximum mass and minimum radius of the neutron star attains their minimum values starting to increase (again) beyond this value. The analysis is complicated by increased numerical oscillations with increasing values of $|\alpha|$. Similarly, we fit the minimum value of NS radius which is attained at the maximum mass with a cubic function
\begin{equation}
R_{\min} = a \alpha_9^3 + b \alpha_9^2 + c \alpha_9 + M_0.
\end{equation}
We find that
$a = 6.18915 \times 10^{-6} \pm 1.625\times 10^{-6}$ km,
$b = 0.00144416  \pm 4.76\times 10^{-5}$ km, 
$c = -0.0469634 \pm 0.0003614 $ km and
$R_0 = 11.2939 \pm 0.003055$ km.

The upper and lower bounds on the value of $\alpha$ presented for each EoS are in the range of $|\alpha| \sim 10^{9}$ cm$^2$. 
Values of $|\alpha|$, that are an order of magnitude smaller than this value, produce results that can not be distinguished from the results obtained within GR. This corresponds to a curvature scale of
$R_0 \sim \alpha^{-1} \sim 10^{-10}$ cm$^{-2}$ and a corresponding
length scale of $L\sim \alpha^{1/2} \sim 10^5$ cm, which is an order
of magnitude smaller than the radius of the neutron star.

%\begin{figure}[H]
%\begin{center}
% \subfigure[]{
% \includegraphics[scale=0.45]{Malpha.eps}
% }
% \subfigure[]{
% \includegraphics[scale=0.45]{Ralpha.eps}
% }
%\end{center}
%\caption{$M_{\max}$ (left panel) and $R_{\min}$ (right panel) changing with $\alpha$ for the polytropic EoS given in Equation~(\ref{polytropic}).}
%\label{fig_alpha}
%\end{figure}

\section{Discussion}

In this work we analyze the neutron star solutions with realistic
EoS' in perturbative $f(R)$ gravity. Among the
modified gravity theories the $f(R)$ theories are relatively simple
to handle. However, even for these theories, the field equations are
complicated and obtaining modified TOV equations in a standard
fashion is difficult. This difficulty is mainly due to field
equations being fourth order unlike in the case of general
relativity, which has second order field equations. In order to
resolve this situation, we adapt a perturbative approach
\cite{Cooney2} in which the extra terms in the gravity action are
multiplied by a `dimensionful' parameter $\alpha$. The extra terms
with some appropriate value of $\alpha$ are supposed to act
perturbatively and modify the results obtained in the case of
general relativity. We present how the perturbative $f(R)$
modifications affect the TOV equations. 

A drawback in using neutron stars for testing alternative theories
of gravity has been the weakly constrained M-R relation. After 
the tight constraints obtained in \cite{OBG10}, it seems this is no longer
quite true. The result of \cite{OBG10} and the measured mass of
PSR J1614-2230 \cite{dem10} excludes many EoS' in the framework of GR.
In the $f(R)=R+\alpha R^2$ gravity model, the value of $\alpha$ provides a new 
degree of freedom and we show in this paper that 
some of the EoS', which are excluded within the framework of GR, 
can now be reconciled with the observations for certain values of $\alpha$.
This then brings the question of degeneracy between using different EoS' and modifying gravity.
In the gravity model studied here, variations in M-R relation comparable to that of using different 
EoS are induced for $\alpha$ being in the order of $10^9$ cm$^2$, which we specify for each EoS. 
An order of magnitude larger values of $\alpha$, which is 
still 5 orders of magnitude smaller than the constraint 
obtained via Gravity Probe B \cite{naf}, does not produce neutron stars with observed properties.
Thus, we argue that $|\alpha| \lesssim 10^{10}$ cm$^2$ is a reasonable constraint independent of
the EoS. We conclude that the presence of uncertainties in the
EoS does not cloak the effect of the free parameter
$\alpha$ on the results. 

For some EoS' (AP4 and GS1) a new stable solution branch ($dM/d\rho_c>0$), 
which does not exist in general relativity, is found. 
This solution branch, for larger values of $|\alpha|$ has a larger domain and joins the conventional 
stable branch beyond some $\alpha$. In this case there is no critical maximum mass to the neutron stars.

One might be curious whether the perturbative approach followed in this paper 
holds for the range of $\alpha$ considered.
For a neutron star the typical value of the Ricci curvature
is calculated to be roughly on the order of $\sim 10^{-12}$
cm$^{-2}$. Therefore in the case of $f(R)=R+\alpha R^2$ model one
easily sees that the perturbative term is $10^{-2}$ orders of
magnitude smaller than the Einstein--Hilbert term $R$, which justifies our approach.

Although we consider 6 representative EoS' here, the above analysis 
repeated with other realistic EoS' will not alter the
order of magnitude of the constraint on $\alpha$ for the following reason: the constraints we
obtained implies a length scale of $R_0^{-1/2}\sim 10^5$ cm which is
only an order of magnitude smaller than the typical radius of a
neutron star, the probe used in this test. This is actually the
length scale below which the gravity models used here \emph{should}
induce modifications on the M-R relation of an object of
size 10 km. This implies that real deviations from general
relativity should be hidden at even much smaller values of
$|\alpha|$.

\acknowledgments

We thank F. \"{O}zel and D. Psaltis for useful discussions and
valuable comments, and J. Lattimer for providing the EoS data. 
This work is supported by the Turkish Council of
Research and Technology (T\"{U}B\.{I}TAK) through grant number
108T686.

\end{document}